\documentclass[journal,twocolumn]{IEEEtran}
\usepackage{setspace}
\usepackage{enumerate}
\usepackage{cite}
\usepackage{graphicx}
\usepackage{epstopdf}
\usepackage{csquotes}
\usepackage{subfig}
\usepackage{mathtools}
\usepackage{amssymb, amsmath,amsthm, amsfonts}
\usepackage{accents}
\usepackage{ifthen}
\usepackage{tikz}
\usepackage{enumerate}
\usepackage{verbatim}
\usepackage{pifont}
\usepackage{kantlipsum}
\allowdisplaybreaks
\usepackage{bm}  
\usepackage{stackengine}
\usepackage{bbm}

\newcommand{\ubar}[1]{\underaccent{\bar}{#1}}
\newcommand{\RN}[1]{%
  \textup{\uppercase\expandafter{\romannumeral#1}}%
}

\DeclareSymbolFont{bbold}{U}{bbold}{m}{n}
\DeclareSymbolFontAlphabet{\mathbbold}{bbold}

\newtheorem{theorem}{Theorem}

\newtheorem{lemma}[theorem]{Lemma}

\newtheorem{definition}[theorem]{Definition}

\newcommand{\ind}{\mathbbm 1}
\newcommand{\Y}{\mathcal{Y}}
\newcommand{\Ucal}{\mathcal{U}}
\newcommand{\W}{\mathcal{W}}
\newcommand{\V}{\mathcal{V}}
\newcommand{\X}{\mathcal{X}}

\newcommand{\Z}{\mathcal{Z}}

\newcommand\blfootnote[1]{%
	\begingroup
	\renewcommand\thefootnote{}\footnote{#1}%
	\addtocounter{footnote}{-1}%
	\endgroup
}

\def\h2{\tilde h}

\def\hm1{\hat h_{-1}}

\begin{document}

\title{Strong Converse for Testing Against  Independence over a Noisy channel}
\author{\IEEEauthorblockN{Sreejith~Sreekumar and
Deniz G\"und\"uz 
}}

\maketitle

\begin{abstract}
A distributed binary hypothesis testing (HT) problem over a noisy (discrete and memoryless) channel studied previously by the authors  is investigated from the perspective of the \textit{strong converse} property. It was shown by Ahlswede and Csisz\'{a}r that a strong converse holds in the above setting when the channel is rate-limited and noiseless. Motivated by this observation, we show that the strong converse continues to hold in the noisy  channel setting for a special case of HT known as testing against independence (TAI). The proof utilizes  the blowing up lemma and the recent change of measure technique of Tyagi and Watanabe as the key tools. 
\end{abstract}

\section{Introduction}
\blfootnote{This work is supported in part by the European Research Council (ERC) through Starting Grant BEACON (agreement \#677854). S. Sreekumar was with the Dept. of Electrical and Electronic Engineering, Imperial College London, at the time of this work. He is now with the Dept.  of Electrical and Computer Engineering, Cornell University, Ithaca, NY 14850, USA (email: sreejithsreekumar@cornell.edu). D. G\"{u}nd\"{u}z is with the Dept. of Electrical and Electronic Engineering, Imperial College London, London SW72AZ, UK (e-mail: d.gunduz@imperial.ac.uk).}
In their seminal paper \cite{Ahlswede-Csiszar}, Ahlswede and Csisz\'{a}r studied a distributed binary hypothesis testing (HT) problem for the joint probability distribution of two correlated discrete memoryless sources. In their setting, one of the sources, denoted by $V$, is observed directly at the \textit{detector}, which performs the test, and the other, denoted by $U$, needs to be communicated to the detector from a remote node, referred to as the \textit{observer}, over a noiseless channel with a transmission rate constraint. 
Given that $n$ independently drawn samples are available at the respective nodes, the two hypotheses are represented using the following null and alternate hypotheses
\begin{subequations} \label{HTeqn}
    \begin{equation}
       H_0:~(U^n,V^n) \sim \prod_{i=1}^n P_{UV},  
    \end{equation}
        \begin{equation}
 H_1:~(U^n,V^n) \sim \prod_{i=1}^n Q_{UV}, 
    \end{equation}
\end{subequations}
where $P_{UV}$ and $Q_{UV}$ denote two arbitrary joint probability distributions over a finite alphabet\footnote{The alphabets $\Ucal$ and $\V$ are assumed to be finite throughout this paper.} $\Ucal \times \V$ ($|\Ucal|,|\V|<\infty$).
The objective is to study the trade-off between the transmission rate,  and the type I and type II error probabilities in HT. This problem  has been  extensively studied thereafter \cite{Han, HK-1989, AH-1989,Shalaby-pap, Shimokawa, Han-Amari-1998, Rahman-Wagner,Watanabe-18,WKJ-2017,SD_2020,Sadaf-Wigger-HTN,WKW_isit19,SD-errexp-2019}. Also, several interesting variants of the basic problem have been considered which includes extensions to multi-terminal settings \cite{Zhao-Lai,Wigger-Timo,Sadaf-Wigger-Li,Cao-Zhou-Tan-2019,Zaidi-Aguerri}, HT under security or privacy constraints \cite{Maggie-Pablo,SD-10-security,SCG-2019,GSSV-2018}, HT with lossy compression \cite{Katz-estdetjourn}, HT in interactive settings \cite{Xiang-Kim-1,Xiang-Kim-2, Katz-collab}, HT with successive refinement \cite{Tian-Chen-2008}, to name a few.

\begin{figure}[t]
\centering
\includegraphics[trim=0cm 0cm 0cm 0cm, clip, width= 0.5\textwidth]{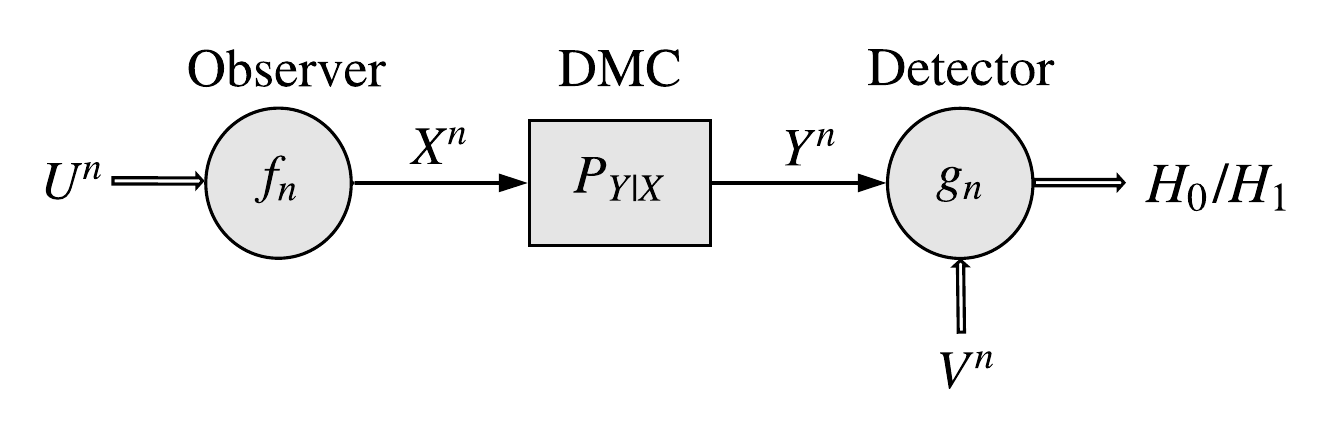}
\caption{Distributed HT over a DMC.} \label{htnoisymodel}
\end{figure}

In this paper, we revisit a distributed HT over a noisy channel problem shown in Fig. \ref{htnoisymodel}, which has been  considered previously in \cite{SD_2020}. 
This problem is a generalization of
\cite{Ahlswede-Csiszar}, in which  
the communication from the observer to the detector happens over a discrete memoryless channel (DMC) with \textit{finite} input and output alphabets, denoted by $\mathcal{X}$ and $\mathcal{Y}$, respectively. Representing the transition probability matrix of the DMC by $P_{Y|X}$, the channel output $Y^n$ given the input $X^n=x^n$ is generated according to the probability law $P_{Y^n|X^n}(y^n|x^n)=\prod_{i=1}^n P_{Y|X}(y_i|x_i)$. 
The observer encodes its observations $U^n$  according\footnote{In \cite{SD_2020}, we allow bandwidth mismatch, i.e., the encoder map is given by  $f_{k,n}: \Ucal^k \mapsto \X^n$, where $k$ and $n$ are positive integers satisfying  $n \leq \tau k$ for some fixed  $\tau \in \mathbb{R}_{\geq 0}$. Here, we consider the special case $k=n$ ($\tau=1$) for simplicity of notation. However, our results extend to any $\tau\in \mathbb{R}_{\geq 0}$ straightforwardly.} to a stochastic map $f_n: \Ucal^n \mapsto \mathcal{P}(\X^n)$, where $\mathcal{P}(\X^n)$ denotes the set of all probability distributions over $\X^n$. The detector outputs the decision  $\hat H=g_n(Y^n,V^n)$ according to a stochastic map $g_n:\Y^n \times \V^n \mapsto \mathcal{P}(\hat{\mathcal{H}})$, where $\mathcal{H}:=\{0,1\}$ and $\mathcal{P}(\hat{\mathcal{H}})$ denotes the set of all probability distributions over support $\mathcal{H}$.  Denoting the true hypothesis as the random variable (r.v.) $H$, the type I and type II error probability for a given encoder-decoder pair $\left(f_n,g_n\right)$  are given by 
\begin{flalign}
    \alpha_n \left(f_n,g_n\right)&=\mathbb{P}\left(\hat H=1|H=0 \right) \notag \\
    &= \mathbb{P}\left(g_n(Y^n,V^n)=1|H=0 \right), \label{t1err} 
    \end{flalign}
and 
\begin{flalign}
    \beta_n \left(f_n,g_n\right)&=\mathbb{P}\left(\hat H=0|H=1 \right) \notag \\
    &= \mathbb{P}\left(g_n(Y^n,V^n)=0|H=1 \right), \label{t2err} 
\end{flalign}
respectively. In \cite{Ahlswede-Csiszar} and \cite{SD_2020}, the goal is to obtain a computable characterization of the optimal \textit{type II error exponent} (henceforth referred to as the error-exponent), i.e., the maximum asymptotic value of the exponent of the type II error probability,  for a fixed non-zero constraint, $\epsilon \in (0,1)$, on the type I error probability. We next define the trade-off studied in \cite{SD_2020} more precisely. 
\begin{definition} \label{defexpdistach} 
An error-exponent $\kappa$ is $\epsilon$-achievable if there exists a sequence of encoding functions $\{f_n\}_{n \in \mathbb{Z}^+}$ and decision rules $\{g_n\}_{n \in \mathbb{Z}^+}$  such that  
\begin{subequations} \label{seqkappa}
\begin{equation}  
  \liminf_{n \rightarrow \infty}  \frac{-1}{n}\log \left( \beta_n \left(f_n, g_n \right) \right)  \geq \kappa,  
\end{equation}
\begin{equation} \label{t1errprbconst}
   \mbox{and }  \alpha_n\left(f_n, g_n \right) \leq \epsilon.  
\end{equation}
\end{subequations} 
\end{definition}
For $\epsilon \in  [0,1]$, let
\begin{align} 
\kappa (\epsilon) & :=  \sup \{ \kappa': \kappa' \mbox{ is } \epsilon\mbox{-achievable} \}. \notag
 \end{align}
 It is well known that since the quantity of interest is the type II error-exponent,  $g_n$ can be restricted to be a deterministic map without any loss of generality (see \cite[Lemma 3]{SCG-2019}).
The decision rule can then be represented as   $g_n(y^n,v^n)=\ind\left((y^n,v^n) \in \mathcal{A}_n\right)$ for some $\mathcal{A}_n \subseteq \Y^n \times \V^n$, where $\ind(\cdot)$ denotes the indicator function.
  
 It is shown in \cite{SD_2020} that $ \lim_{\epsilon \rightarrow 0} \kappa(\epsilon)$ has an exact single-letter characterization for the special case known as \textit{testing against independence} (TAI), in which, $Q_{UV}$ factors as a product of marginals of $P_{UV}$, i.e., $Q_{UV}:=P_U \times P_V$. To state the result, let $C:=C(P_{Y|X})$ denote the capacity of the channel $P_{Y|X}$, and let
\begin{align}
    \theta (P_{UV},C):= \sup \left\{
  \begin{aligned}
  I(V;W):~& \exists~ W \mbox{ s.t. } I(U;W) \leq C,\\&V-U-W.
  \end{aligned}
\right\}. \label{expcharsing}
\end{align}
It is proved in \cite[Proposition 7]{SD_2020} that
\begin{align}
    \lim_{\epsilon \rightarrow 0} \kappa(\epsilon)=  \theta (P_{UV},C).\notag
\end{align}
In this paper, we show the strong converse for the above result, namely, that
\begin{align}
    \kappa(\epsilon)=  \theta (P_{UV},C),~\forall~\epsilon \in (0,1).\notag
\end{align}
This result completes the characterization of $\kappa(\epsilon)$ in terms of $\theta$ for all values of $\epsilon$, and extends the strong converse result proved in \cite[Proposition 2]{Ahlswede-Csiszar} for the special case of rate-limited noiseless channels. However, it is to be noted that while the strong converse proved in \cite{Ahlswede-Csiszar} holds for all hypothesis tests given in \eqref{HTeqn} such that $Q_{UV}(u,v)>0, ~\forall~(u,v) \in \Ucal \times \V$, our result is limited to TAI. 

Before delving into the proof, we briefly describe the technique and tools used in \cite{Ahlswede-Csiszar} to prove the strong converse, and highlight the challenges of extending their proof to the noisy channel setting. 
The key tools used to prove \cite[Proposition 2]{Ahlswede-Csiszar} are the so-called \textit{blowing-up lemma} \cite{AGK-1976} and a \textit{covering lemma} \cite{Ahlswede-Csiszar}. However, it can be seen from the proof therein that the application of the \textit{covering lemma} to prove the strong converse relies crucially on the fact that the channel from the encoder to the detector is noiseless (i.e. deterministic). Thus, it is not possible to directly follow their technique in our noisy channel setting and arrive at the strong converse result. Alternatively, we will use a \textit{change of measure} technique introduced in \cite{Tyagi-Watanabe}, in conjunction with the blowing-up lemma for this purpose. 

The change of measure technique by itself does not appear sufficient for proving a strong converse in our setting. This is so because a critical aspect for the technique to work is to find a (decoding) set $\mathcal{B}_n \subseteq \Ucal^n \times \V^n $ of non-vanishing probability (with respect to $n$) under the null hypothesis such that for a given $\mathcal{A}_n \subseteq \Y^n \times \V^n$ satisfying the type I error probability constraint and each $(u^n,v^n) \in \mathcal{B}_n$, $(Y^n_{|u^n},v^n) \in \mathcal{A}_n$ with probability one (or tending to one with $n$), where $ Y^n_{|u^n} \sim P_{Y^n|U^n}(\cdot|u^n)$. Note that in the noiseless channel case, the set $\mathcal{B}_n$ satisfying the above conditions can be obtained by simply taking
\begin{align}
 \mathcal{B}_n:=\{(u^n,v^n): \big(f_n(u^n),v^n\big) \in \mathcal{A}_n\},\notag
\end{align}
as is done in \cite{Cao-Zhou-Tan-2019} for a deterministic $f_n$. However, this is no longer possible when the channel is noisy. To tackle this issue, we first obtain a set $\mathcal{B}_n$ of sufficiently large probability under the null hypothesis such that for each $(u^n,v^n) \in \mathcal{B}_n$, $(Y^n_{|u^n},v^n) \in \mathcal{A}_n$ with a positive probability bounded away from zero. The blowing-up lemma then guarantees that it is possible to obtain a modified decision region $\bar{\mathcal{A}}_n$ such that uniformly for each $(u^n,v^n) \in \mathcal{B}_n$, $(Y^n_{|u^n},v^n) \in \mathcal{A}_n$ with an overwhelmingly large probability. This enables us to prove the strong converse in our setting via the technique in \cite{Tyagi-Watanabe}.

We next state a  non-asymptotic version of the blowing up lemma given in \cite{Marton-96}, which will be used in the proof of Theorem \ref{strongconv-main} below.
For any set $\mathcal{D} \subset \Z^n$, let $\Gamma^{l}(\mathcal{D})$ denote the Hamming $l-$neighbourhood of $\mathcal{D}$, i.e.,
\begin{flalign}
   \Gamma^{l}(\mathcal{D}) &:=\{\tilde z^n \in \Z^n: d_H(z^n,\tilde z^n) \leq l \mbox{ for some } z^n \in \mathcal{D}\},\label{hamnb}
\end{flalign} 
where 
\begin{align}
   d_H(z^n,\tilde z^n)=\sum_{i=1}^n \ind(z_i \neq \tilde z_i). \notag
\end{align}
 
\begin{lemma}\cite{Marton-96} \label{blowinguplemma}
Let $Z_1,\ldots,Z_n$ be $n$ independent r.v.'s taking values in a finite set $\Z$. Then, for any set $\mathcal{D} \subseteq \Z^n$ with $P_{Z^n}(\mathcal{D})>0$,
\begin{align}
  P_{Z^n}(\Gamma^l(\mathcal{D})) &\geq 1-e^{\left[-\frac 2n \left(l-\sqrt{\frac n2 \log\left(\frac{1}{P_{Z^n(\mathcal{D})}}\right)}\right)^2\right]},\notag \\
  & \qquad \forall~ l> \sqrt{\frac{n}{2} \log \left(\frac{1}{P_{Z^n}(\mathcal{D})} \right)}. \notag
\end{align}
\end{lemma}
The next lemma provides a characterization of $\theta(P_{UV},C)$ in terms of hyper-planes in the error exponent-capacity region. 
\begin{lemma} \label{lemmsingletchar}
\begin{align}
\theta(P_{UV},C) &=\inf_{\mu>0} \theta_{\mu}(P_{UV},C),\notag
\end{align}
where 
\begin{align}
 \theta_{\mu}(P_{UV},C):=\sup_{\substack{P_{W|U}:\\
 V-U-W}} I(V;W)+\mu (C-I(U;W)).   \label{auxmaxmuval}
\end{align}
\end{lemma}
\begin{IEEEproof}
Let 
\begin{align}  
& \mathcal{R}:=\left\lbrace\begin{aligned}
(\theta,C) \in \mathbb{R}_{\geq 0}^2:&~ \exists~ W \mbox{ s.t. } V-U-W, ~ \theta \leq I(V;W)  \\
&~\mbox{ and } I(U;W) \leq C\end{aligned}\right\rbrace. \label{chartwodim} 
\end{align}
By the Fenchel-Eggleston-Caratheodory theorem \cite{Elgamalkim}, it is sufficient to take $|\W| \leq |\Ucal|+1$ in \eqref{chartwodim}. Hence, noting that $\mathcal{R}$ is a closed convex set, it can be represented via the intersection of half spaces as
\begin{align}
  \mathcal{R}:=\bigcap_{\mu>0}\{(\theta,C): \theta-\mu C \leq R_{\mu} \}, \notag 
\end{align}
where 
\begin{align}
    R_{\mu}:= \max_{\substack{W:\\W-U-V}} I(V;W)-\mu I(U;W).\notag
\end{align}
This implies that
\begin{align}
   \theta(P_{UV},C)&:=\sup \{\theta: (\theta,C) \in \mathcal{R}\}\notag\\
  & =\inf_{\mu>0} R_{\mu}+\mu C \notag\\
   &=\inf_{\mu>0}\theta_{\mu}(P_{UV},C).\notag
\end{align}
\end{IEEEproof}
\section{Main result}
The main result of the paper is stated next.
\begin{theorem} \label{strongconv-main}
\begin{align}
  \kappa(\epsilon)=  \theta (P_{UV},C),~\forall~\epsilon \in (0,1).\notag
\end{align}
\end{theorem}
\begin{IEEEproof}Let $f_n$ and $g_n$ denote an encoder-decoder pair specified by $P_{X^n|U^n}$ and $\mathcal{A}_n$, respectively, that satisfies \eqref{t1errprbconst}.\\[5 pt]
\underline{\textbf{Constructing reliable decision regions $\bar{\mathcal{B}}_n$ and $\bar{\mathcal{A}}_n$}:}\\[10 pt]
 Note that $\mathcal{A}_n$ can be written in the form
\begin{align}
\mathcal{A}_n= \bigcup\limits_{v^n \in \V^n} \mathcal{A}(v^n) \times \{v^n\},
\end{align}
where $\mathcal{A}(v^n):=\{y^n \in \Y^n: (y^n,v^n) \in \mathcal{A}_n\}$. \\
Let 
\begin{flalign}
    \mathcal{B}_n(\gamma):=\left\lbrace
                \begin{aligned} \notag
                  (u^n,v^n,x^n):~ & P_{X^n|U^n}(x^n|u^n)>0, \\
& P_{Y^n|X^n}\left(\mathcal{A}(v^n)|x^n\right) \geq \gamma
                \end{aligned}
             \right\rbrace.&&
\end{flalign}
Then, it follows from \eqref{t1errprbconst} that for sufficiently large $n$,
\begin{align}
    \mathbb{P}\left((U^n,V^n,X^n) \in  \mathcal{B}_n(\gamma) \big|H=0\right) \geq \frac{1-\epsilon-\gamma}{1-\gamma}. \notag
\end{align}
Selecting $\gamma=\frac{1-\epsilon}{2}$ yields
\begin{align}
     \mathbb{P}\left((U^n,V^n,X^n) \in  \mathcal{B}_n(\gamma) \big|H=0\right) \geq \frac{1-\epsilon}{1+\epsilon}. \label{probsetlb}
\end{align}
Let  
\begin{align}
\bar{\mathcal{B}}_n&:= \mathcal{B}_n\left(\frac{1-\epsilon}{2}\right), \notag\\
\mathcal{B}_{v^n}&:=\{(u^n,x^n): (u^n,v^n,x^n) \in \bar{\mathcal{B}}_n\}, \notag \\
\hat{\mathcal{B}}_n&:=\{(v^n,x^n):~ (u^n,v^n,x^n) \in \bar{\mathcal{B}}_n \mbox{ for some } u^n \in \Ucal^n  \}, \notag \\
    l_n&:=\left\lceil \frac{1}{\sqrt{2}} \left( \sqrt{nb(n)}+\sqrt{n\log\left(\frac{1+\epsilon}{1-\epsilon}\right)}\right)\right\rceil, \notag \\
    \bar{\mathcal{A}}(v^n)&:=\Gamma^{l_n}(\mathcal{A}(v^n)), \notag
\end{align}
where $b:\mathbb{N} \mapsto \mathbb{R}_{\geq 0}$ is a function (that will be optimized later) such that $\lim_{n \rightarrow \infty} b(n)=\infty$. It follows from Lemma \ref{blowinguplemma} that 
\begin{align}
    P_{Y^n|X^n}(\bar{\mathcal{A}}(v^n)|x^n) & \geq \epsilon_n':=1-e^{-b(n)}  \xrightarrow{(n)}1, \label{type1probtend1} 
\end{align}
for every $(v^n,x^n) \in \hat{\mathcal{B}}_n$, since 
\begin{align}
    P_{Y^n|X^n}\left(\mathcal{A}(v^n)|x^n\right) \geq 0.5(1-\epsilon). \notag
\end{align} Let
\begin{flalign}
    \ubar p:=\min \left\{P_{Y|X}(y|x), (x,y) \in \X \times \Y:P_{Y|X}(y|x)>0\right\}.\notag&&
\end{flalign}
Then, for any $(v^n,x^n) \in \hat{\mathcal{B}}_n$,  we can write using  \eqref{hamnb}  that 
\begin{flalign}
    & P_{Y^n|X^n}(\bar{\mathcal{A}}(v^n)|x^n) \notag \\
    &  \leq \sum_{y^n \in \mathcal{A}(v^n)}~\sum_{\tilde y^n \in \Gamma^{l_n}(y^n)} P_{Y^n|X^n}(\tilde y^n|x^n) \notag \\
    &= \sum_{y^n \in \mathcal{A}(v^n)}~\sum_{\tilde y^n \in \Gamma^{l_n}(y^n)} P_{Y^n|X^n}(\tilde y^n|x^n)~\ind_{\left\{P_{Y^n|X^n}(\tilde y^n|x^n) >0\right\}}  \notag \\
    & \leq \sum_{y^n \in \mathcal{A}(v^n)}~\sum_{\tilde y^n \in \Gamma^{l_n}(y^n)} P_{Y^n|X^n}(y^n|x^n) \ubar p^{-l_n}\label{blowingupseq} \\
    &\leq \sum_{y^n \in \mathcal{A}(v^n)} P_{Y^n|X^n}(y^n|x^n)~\binom{n}{l_n}~ |\Y|^{l_n}\ubar p^{-l_n} \notag \\
    & \leq (|\Y|ne)^{l_n}(\ubar p~l_n)^{-l_n}P_{Y^n|X^n}(\mathcal{A}(v^n)|x^n), \label{t2errbnd}&&
\end{flalign}
where, \eqref{blowingupseq} follows since for each $y^n \in \mathcal{A}(v^n)$ and $\tilde y^n \in \Gamma^{l_n}(y^n)$,
\begin{align}
    P_{Y^n|X^n}(\tilde y^n|x^n)~\ubar p^{l_n} \leq P_{Y^n|X^n}(y^n|x^n),\notag 
\end{align} 
and \eqref{t2errbnd}
is due to the inequality $\binom{n}{l_n} \leq \left(\frac{ne}{l_n}\right)^{l_n}$.

Let the new decision rule $\bar g_n$ be given by $\bar g_n(y^n,v^n)= \ind\left((y^n,v^n) \in  \bar{\mathcal{A}}_n\right)$, where 
\begin{align}
   \bar{\mathcal{A}}_n:=\bigcup_{v^n \in \V^n} \bar{\mathcal{A}}(v^n) \times \{v^n\}.\notag
\end{align}
Note that it follows from \eqref{t2errbnd} that
\begin{align}
    \beta_n\left(f_n,\bar g_n\right) \leq  \beta_n\left(f_n, g_n\right) \left(\frac{|\Y|ne}{\ubar pl_n}\right)^{l_n}. \label{uppbndt2errp}
\end{align}
\underline{\textbf{Change of measure via  construction of a truncated }}\\[5 pt] \underline{\textbf{distribution:}}\\[5 pt]
We now use the change of measure technique in \cite{Tyagi-Watanabe} by considering the new decision rule $\bar g_n$ (with acceptance region $\bar{\mathcal{A}}_n$ for $H_0$) to prove the strong converse. To that purpose, define a new truncated distribution 
\begin{flalign}
   & P_{\tilde U^n \tilde V^n \tilde X^n \tilde Y^n}(u^n,v^n,x^n,y^n) \notag \\
   &:= \frac{P_{U^nV^n}(u^n,v^n)P_{X^n|U^n}(x^n|u^n) }{P_{U^nV^nX^n}(\bar{\mathcal{B}}_n)}~\ind((u^n,v^n,x^n) \in \bar{\mathcal{B}}_n)\notag \\
   &\qquad P_{Y^n|X^n}(y^n|x^n). \label{truncdist}&&
\end{flalign}
\underline{\textbf{Bounding type II error-exponent via the weak converse:}}\\[5 pt]
From  \eqref{type1probtend1} and \eqref{truncdist}, note that the type I error probability for  the hypothesis test between distributions $P_{\tilde U^n \tilde V^n}$ and $P_{U^n} \times P_{V^n}$ (under the null and alternate hypotheses, respectively), channel input $\tilde X^n=f_n(\tilde U^n)$, and decision rule $\bar g_n$ tends to zero asymptotically as $e^{-b(n)}$.
Then, by the weak converse for HT based on the data processing inequality for KL divergence (see \cite{Ahlswede-Csiszar}, \cite{SD_2020}), it follows that
 \begin{flalign}
     & - \log\left(\beta_n(f_n,\bar g_n)\right) \notag \\
     &\leq \frac{1}{\epsilon_n'}\left(D(P_{\tilde V^n \tilde Y^n}||P_{V^n} \times P_{Y^n}) +\log 2\right). \label{weakconvex} &&
 \end{flalign}
Next, note that for $v^n$ such that $|\mathcal{B}_{v^n}| \geq 1$, we have
\begin{flalign}
    P_{\tilde V^n}(v^n)&=\sum_{(u^n,x^n) \in \mathcal{B}_{v^n}} P_{\tilde U^n \tilde V^n \tilde X^n}(u^n, v^n,x^n) \notag \\
    &= \frac{1}{P_{\tilde U^n \tilde V^n\tilde X^n}(\bar{\mathcal{B}}_n)}\sum_{(u^n,x^n) \in \mathcal{B}_{v^n}} P_{U^nV^nX^n}(u^n, v^n) \notag \\
    &\leq \frac{ P_{V^n}(v^n)}{P_{\tilde U^n \tilde V^n \tilde X^n}(\bar{\mathcal{B}}_n)} \leq \frac{1+\epsilon}{1-\epsilon}P_{V^n}(v^n). \label{bndmargv} &&
\end{flalign}
Similarly, for all $y^n \in \Y^n$, we have 
\begin{align}
    P_{\tilde Y^n}(y^n)& \leq \frac{ P_{Y^n}(y^n)}{P_{\tilde U^n \tilde V^n \tilde X^n}(\bar{\mathcal{B}}_n)} \leq \frac{1+\epsilon}{1-\epsilon}P_{Y^n}(y^n). \label{bndmargu}
\end{align}
Substituting  \eqref{bndmargv} and \eqref{bndmargu} in \eqref{weakconvex} yields 
\begin{flalign}
   &- \log\Big(\beta_n(f_n,\bar g_n)\Big) \notag \\
   &\leq  \frac{1}{\epsilon_n'} \bigg(D\left(P_{\tilde V^n \tilde Y^n}||P_{\tilde V^n} \times P_{\tilde Y^n}\right)+2\log\bigg(\frac{1+\epsilon}{1-\epsilon}\bigg)\notag \\
   & \qquad \qquad + \log 2\bigg) \notag \\
   &= \frac{1}{\epsilon_n'} \bigg(I(\tilde V^n;\tilde Y^n)+2\log\bigg(\frac{1+\epsilon}{1-\epsilon}\bigg)+ \log 2\bigg).&& \label{weakconveq}
\end{flalign}
Combining \eqref{weakconveq} with \eqref{uppbndt2errp}, we obtain that
\begin{flalign}
& - \log\left(\beta_n\left(f_n,g_n\right)\right) \notag \\
&\leq  \frac{1}{\epsilon_n'} \left(I(\tilde V^n;\tilde Y^n)+2\log\left(\frac{1+\epsilon}{1-\epsilon}\right)+ \log 2 \right) \notag \\
&\qquad \qquad +l_n \log\left(\frac{|\Y|ne}{\ubar p l_n}\right) \notag \\
&:=\zeta_n+\frac{1}{\epsilon_n'}I(\tilde V^n;\tilde Y^n). & \notag
\end{flalign}
Now, notice from \eqref{probsetlb} and \eqref{truncdist} that
\begin{flalign}
D(P_{\tilde U^n }||P_{U^n}) &\leq D(P_{\tilde U^n \tilde V^n\tilde X^n \tilde Y^n}||P_{U^nV^nX^nY^n}) \label{applydpikl}\\
&=D(P_{\tilde U^n \tilde V^n \tilde X^n}||P_{U^nV^nX^n}) \notag \\
&=\log\left(\frac{1}{P_{U^nV^n X^n}(\bar{\mathcal{B}}_n)}\right) \notag \\
&\leq \log\left(\frac{1+\epsilon}{1-\epsilon}\right), \label{bnddivcost}&&
\end{flalign}
where \eqref{applydpikl} follows from the log-sum inequality \cite{Csiszar-Korner}. Also, observe from \eqref{truncdist} that the Markov chain $\tilde V^n-\tilde U^n-\tilde X^n-\tilde Y^n$ holds under $ P_{\tilde U^n \tilde V^n \tilde X^n \tilde Y^n}$, and that $P_{\tilde Y^n|\tilde X^n}(y^n|x^n)=\prod_{i=1}^n P_{Y|X}(y_i|x_i)$. From this, it follows via the data processing inequality that
\begin{align}
    I(\tilde U^n;\tilde Y^n) \leq I(\tilde X^n;\tilde Y^n) \leq nC.\notag
\end{align}
Thus, we have for any $\mu \geq 0,~\nu \geq 0$ that
\begin{flalign}
 &- \epsilon_n' \log\left(\beta_n\left(f_n,g_n\right)\right) \notag \\
 &\leq    I(\tilde V^n;\tilde Y^n)+ n\mu C- \mu I(\tilde U^n;\tilde Y^n)+\epsilon_n' \zeta_n \notag\\
 & \leq  I(\tilde V^n;\tilde Y^n)+n\mu C- \mu I(\tilde U^n;\tilde Y^n)+\epsilon_n' \zeta_n  \notag \\
 &\quad -\nu I(\tilde V^n;\tilde Y^n|\tilde U^n) -(\nu+\mu) D(P_{\tilde U^n\tilde V^n}||P_{U^nV^n}) \notag \\
 & \qquad +(\nu+\mu)\log\left(\frac{1+\epsilon}{1-\epsilon}\right)\label{adddivcost} \\
 &=R_{\mu,\nu}^{(n)}+(\nu+\mu)\log\left(\frac{1+\epsilon}{1-\epsilon}\right)+\epsilon_n' \zeta_n, \label{outbndnlet} &&
\end{flalign}
where 
\begin{align}
  R_{\mu,\nu}^{(n)}&:= I(\tilde V^n;\tilde Y^n) +n\mu C- \mu\big( I(\tilde U^n;\tilde Y^n) \notag \\
  &\quad+ D(P_{\tilde U^n\tilde V^n}||P_{U^nV^n})\big)-\nu\big( I(\tilde V^n;\tilde Y^n|\tilde U^n) \notag \\
  &~~\quad + D(P_{\tilde U^n\tilde V^n}||P_{U^nV^n})\big). \label{multiletchar}
\end{align}
Equation \eqref{adddivcost} follows from \eqref{bnddivcost} and the fact that $I(\tilde V^n;\tilde Y^n|\tilde U^n)=0$ (which in turn holds due to the Markov chain $\tilde V^n-\tilde U^n-\tilde Y^n$ under distribution $P_{\tilde U^n \tilde V^n \tilde Y^n}$).\\[5 pt] 
\underline{\textbf{Single-letterization of  $ R_{\mu,\nu}^{(n)}$ and applying Lemma \ref{lemmsingletchar}}}\\[5 pt]
We will show in Appendix \ref{appsinglett} that $ R_{\mu,\nu}^{(n)}$ single-letterizes, i.e., 
\begin{align}
    R_{\mu,\nu}^{(n)} \leq n R^s_{\mu,\nu}(P_{UV},C), \label{subaddit}
\end{align}
where 
\begin{align}
    & R^s_{\mu,\nu}(P_{UV},C) \notag \\
    &:=\sup_{\substack{P_{\tilde U \tilde V \tilde W}\\ \in \mathcal{P}_{\Ucal \V \tilde \W}} } \Big[ I(\tilde V;\tilde W)+\mu C-\mu I(\tilde U;\tilde W)-(\nu+\mu) I(\tilde V;\tilde W|\tilde U) \notag \\
    &\qquad \qquad ~~-(\nu+\mu)D(P_{\tilde U \tilde V}||P_{UV}) \Big]. \label{singletvalmax}
\end{align}
By the Fenchel-Eggleston-Caratheodory theorem \cite{Elgamalkim}, $|\tilde{\W}|$ can be restricted to be finite (a function of $|\Ucal|$ and $|\V|$) in the maximization in \eqref{singletvalmax}. Thus, the supremum in \eqref{singletvalmax} is actually a maximum. Assuming \eqref{subaddit} holds, we can write from \eqref{outbndnlet} that
\begin{flalign}
   & - \epsilon_n' \log\left(\beta_n\left(f_n,g_n\right)\right) \notag \\
    &\leq n R^s_{\mu,\nu}(P_{UV},C)+(\nu+\mu)\log\left(\frac{1+\epsilon}{1-\epsilon}\right)+\epsilon_n' \zeta_n. \label{finoutbnd} &&
\end{flalign}
For a given $\mu$, $\nu$, let $P_{U_{\mu,\nu}V_{\mu,\nu}W_{\mu,\nu}}$ achieve the maximum in \eqref{singletvalmax}. Then, we can write for $P_{UVW_{\mu,\nu}}:=P_{UV}P_{W_{\mu,\nu}|U}:=P_{UV}P_{W_{\mu,\nu}|U_{\mu,\nu}}$ that
\begin{flalign}
 & R^s_{\mu,\nu}(P_{UV},C) \notag\\
& = I(V_{\mu,\nu};W_{\mu,\nu})+\mu C-\mu I(U_{\mu,\nu};W_{\mu,\nu}) \notag \\
&~ \qquad-(\nu+\mu) I(V_{\mu,\nu};W_{\mu,\nu}|U_{\mu,\nu}) \notag \\& \qquad -(\nu+\mu)D(P_{U_{\mu,\nu}V_{\mu,\nu}}||P_{UV}) \label{achmaxrmunu}\\
 &\leq I(V_{\mu,\nu};W_{\mu,\nu})+\mu C-\mu I(U_{\mu,\nu};W_{\mu,\nu}) \notag\\
 & \leq I(V;W_{\mu,\nu})+\mu C-\mu I(U;W_{\mu,\nu}) \notag\\
 & \qquad + |I(V_{\mu,\nu};W_{\mu,\nu})- I(V;W_{\mu,\nu})|\notag \\
 & \qquad + \mu|I(U_{\mu,\nu};W_{\mu,\nu})- I(U;W_{\mu,\nu})|\notag\\
 & \leq \theta_{\mu}(P_{UV},C)+ |I(V_{\mu,\nu};W_{\mu,\nu})- I(V;W_{\mu,\nu})| \notag\\
 &\qquad +\mu |I(U_{\mu,\nu};W_{\mu,\nu})- I(U;W_{\mu,\nu})|. \label{bndrateregext} 
\end{flalign}
We next upper bound the second and third terms in \eqref{bndrateregext} similar in spirit to \cite{Cao-Zhou-Tan-2019}.  Note that 
\begin{flalign}
&R^s_{\mu,\nu}(P_{UV},C) \notag \\
& \geq \inf_{\mu>0,\nu>0} R^s_{\mu,\nu}(P_{UV},C) \notag \\ 
& \geq \theta(P_{UV},C) \geq I(V;W_{\mu,\nu})+\mu C-\mu I(U;W_{\mu,\nu}). \label{bndcompratereg} &&
\end{flalign}
Then, we can write that
\begin{flalign}
& \nu D(P_{U_{\mu,\nu}V_{\mu,\nu}W_{\mu,\nu}}||P_{UVW_{\mu,\nu}}) \notag \\
    &=\nu \left(I(V_{\mu,\nu};W_{\mu,\nu}|U_{\mu,\nu})+D(P_{U_{\mu,\nu}V_{\mu,\nu}}||P_{UV}) \right) \notag \\
    &\leq |I(V_{\mu,\nu};W_{\mu,\nu})-I(V;W_{\mu,\nu})| \notag \\
    &\qquad +\mu |I(U_{\mu,\nu};W_{\mu,\nu})-I(U;W_{\mu,\nu})| \label{applybndmutterms} \\
    &\leq \log(|V||W_{\mu,\nu}|)+\mu \log(|U||W_{\mu,\nu}|):=\chi(\mu), \label{applybnd} &&
\end{flalign}
where we used  \eqref{achmaxrmunu} and \eqref{bndcompratereg} to obtain \eqref{applybndmutterms}. Thus, we have
\begin{align}
  D(P_{U_{\mu,\nu}V_{\mu,\nu}W_{\mu,\nu}}||P_{UVW_{\mu,\nu}})  \leq \frac{\chi(\mu)}{\nu}. \label{bnddivgtot}
\end{align}
Denoting the total variation distance between distributions $P_{V_{\mu,\nu}W_{\mu,\nu}}$ and $P_{VW_{\mu,\nu}}$ by 
\begin{align}
&d(P_{V_{\mu,\nu}W_{\mu,\nu}},P_{VW_{\mu,\nu}}) \notag \\
&:= \frac{1}{2} \sum_{\substack{(v,w') \\ \in \V \times \W_{\mu,\nu}}} |P_{V_{\mu,\nu}W_{\mu,\nu}}(v,w')-P_{VW_{\mu,\nu}}(v,w')|, \notag
\end{align}
we have by Pinsker's inequality that
\begin{align}
  d(P_{V_{\mu,\nu}W_{\mu,\nu}},P_{VW_{\mu,\nu}}) &\leq \sqrt{\frac{D(P_{V_{\mu,\nu}W_{\mu,\nu}}||P_{VW_{\mu,\nu}})}{2}}  \notag\\
  &\leq  \sqrt{\frac{\chi(\mu)}{2\nu}}.\notag
\end{align}
For $\nu=\Theta(\sqrt{n})$, applying  \cite[Lemma 2.7]{Csiszar-Korner}, we can write 
\begin{flalign}
   & |H(P_{V_{\mu,\nu}W_{\mu,\nu}})-H(P_{VW_{\mu,\nu}})| \notag  \\
   & \leq d(P_{V_{\mu,\nu}W_{\mu,\nu}},P_{VW_{\mu,\nu}}) \log\left(\frac{|V||W_{\mu,\nu}|}{d(P_{V_{\mu,\nu}W_{\mu,\nu}},P_{VW_{\mu,\nu}})}\right) \notag \\
   & \leq  \sqrt{\frac{\chi(\mu)}{2\nu}} \log\left(\frac{|V||W_{\mu,\nu}|}{ \sqrt{\frac{\chi(\mu)}{2\nu}}}\right)=\Theta\left( \sqrt{\frac{\mu}{\nu}}\log\left(\frac{\mu}{\nu}\right)\right). \label{bndentpy1}
\end{flalign}
From \eqref{bndentpy1}, it follows that for $\nu=\Theta(\sqrt{n})$,
\begin{align}
& |I(V_{\mu,\nu};W_{\mu,\nu})- I(V;W_{\mu,\nu})|\leq \Theta\left( \sqrt{\frac{\mu}{\nu}}\log\left(\frac{\mu}{\nu}\right)\right). \label{bndmutinf1} 
\end{align}
Similarly,  using \eqref{bnddivgtot},   we obtain for $\nu=\Theta(\sqrt{n})$ that
\begin{align}
& |I(U_{\mu,\nu};W_{\mu,\nu})- I(U;W_{\mu,\nu})| \leq \Theta\left( \sqrt{\frac{\mu}{\nu}}\log\left(\frac{\mu}{\nu}\right)\right).  \label{bndmutinf2}
\end{align}
Combining \eqref{finoutbnd}, \eqref{bndrateregext}, \eqref{bndmutinf1} and \eqref{bndmutinf2} yields
\begin{flalign}
 &- \epsilon_n' \log\left(\beta_n\left(f_n,g_n\right)\right) \notag \\
 & \leq n \theta_{\mu}(P_{UV},C)+ n \Theta\left( \sqrt{\frac{\mu}{\nu}}\log\left(\frac{\mu}{\nu}\right)\right)\notag \\
 & \qquad +(\nu+\mu)\log\left(\frac{1+\epsilon}{1-\epsilon}\right)+\epsilon_n' \zeta_n. \label{finbnderrexp}&&
\end{flalign}
Since \eqref{finbnderrexp} holds for any $\mu>0$ and  $\nu=\Theta(\sqrt{n})$,  we have
\begin{flalign}
 &- \epsilon_n' \log\left(\beta_n\left(f_n,g_n\right)\right) \notag \\
 &\leq n \theta_{\mu}(P_{UV},C)+  \Theta\left( \sqrt{\mu}n^{\frac 34}\log\left(\frac{\mu}{n}\right)\right)\notag \\
 &\quad +(\sqrt{n}+\mu)\log\left(\frac{1+\epsilon}{1-\epsilon}\right)+\epsilon_n' \zeta_n. \label{finbnderrexp2}&&
 \end{flalign}
By selecting $b(n)=\log n$ in the definition of $ \zeta_n$, dividing by $n$ and taking limit supremum on both sides of \eqref{finbnderrexp2}, we obtain
\begin{flalign}
    &\limsup_{n \rightarrow \infty} \frac{-1}{n} \log\left(\beta_n\left(f_n,g_n\right)\right) \notag \\
    &= \limsup_{n \rightarrow \infty} \frac{-\epsilon_n'}{n} \log\left(\beta_n\left(f_n,g_n\right)\right) \notag \\
    & \leq \theta_{\mu}(P_{UV},C)+\limsup_{n \rightarrow \infty}\frac{\zeta_n}{n}=\theta_{\mu}(P_{UV},C). \notag&&
\end{flalign}
Finally, taking infimum over $\mu>0$ on both sides and noting that $\left(f_n,g_n\right)$ was arbitrary, we establish the strong converse via Lemma \ref{lemmsingletchar}.
\end{IEEEproof}
\section{Conclusion}
In this paper, we have proved the strong converse for distributed hypothesis testing over a noisy channel for the special case of TAI. While we believe that the strong converse should hold in general for distributed HT over a DMC, the proof technique used here appears inadequate for this purpose, as the change of measure technique relies on the availability of a single-letter characterization of the optimal error-exponent for the vanishing type I error probability constraint. A promising technique to that end that we will pursue in the future is the one  based on reverse hypercontractivity  proposed in \cite{Liu-handel-verdu-2017}. 
\begin{appendices}
\section{Single-letterization of  $ R_{\mu,\nu}^{(n)}$} \label{appsinglett}
We prove \eqref{subaddit} below.
Using standard single-letterization steps with the auxiliary r.v. identification $W=(W_Q,Q)$, where $Q$ is a r.v. uniformly distributed in $[1:n]$ and independent of all the other r.v.'s, and $W_i:=(\tilde V^{i-1},\tilde Y^n)$, we can write
\begin{flalign}
    I(\tilde V^n;\tilde Y^n)&=H(\tilde V^n)-H(\tilde V^n|\tilde Y^n) \notag \\
    & \leq \sum\nolimits_{i=1}^n H(\tilde V_i)-\sum\nolimits_{i=1}^n H(\tilde V_i|\tilde V^{i-1}, \tilde Y^n) \notag \\
    & =  n I(\tilde V_Q;W_Q,Q)=n I(\tilde V_Q;W), \label{singletstp1} 
\end{flalign}
and 
\begin{flalign}
    H(\tilde V^n|\tilde U^n,\tilde Y^n)&= \sum\nolimits_{i=1}^n H(\tilde V_i|\tilde V^{i-1}, \tilde U^n, \tilde Y^{n})  \notag \\
    & \leq \sum\nolimits_{i=1}^n H(\tilde V_i|\tilde V^{i-1}, \tilde U_i, \tilde Y^{n}) \notag \\
    & =n H(\tilde V_Q|W_Q,\tilde U_Q,Q)\notag \\
    &=n H(\tilde V_Q|W,\tilde U_Q). \label{singletstp3} &&
\end{flalign}
Also, 
\begin{flalign}
    I(\tilde U^n;\tilde Y^n)&=I(\tilde U^n, \tilde V^n;\tilde Y^n) \label{Markchnapp} \\
    &=H(\tilde U^n, \tilde V^n)-H(\tilde U^n, \tilde V^n|\tilde Y^n), \label{rateconstsinglet} &&
\end{flalign}
where \eqref{Markchnapp} is due to the Markov chain $\tilde V^n-\tilde U^n-\tilde Y^n$ that holds under $P_{\tilde U^n \tilde V^n \tilde Y^n}$. The last term in \eqref{rateconstsinglet} can be single-letterized as follows:
\begin{align}
   H(\tilde U^n, \tilde V^n|\tilde Y^n)&= \sum\nolimits_{i=1}^n H(\tilde U_i,\tilde V_i|\tilde U^{i-1},\tilde V^{i-1}, \tilde Y^n) \notag \\
   & \leq \sum\nolimits_{i=1}^n H(\tilde U_i,\tilde V_i|\tilde V^{i-1}, \tilde Y^n) \notag \\
   &= n H(\tilde U_Q,\tilde V_Q|W_Q,Q) \notag \\
   &=nH(\tilde U_Q,\tilde V_Q|W). \label{singletstp2} &&
\end{align}
Finally, we have the following sequence of steps to single-letterize  the remaining terms in \eqref{multiletchar}:
\begin{flalign}
 & H(\tilde U^n,\tilde V^n)+D(P_{\tilde U^n \tilde V^n}||P_{U^nV^n}) \notag \\
    &= \sum_{(u^n,v^n) \in \Ucal^n \times \V^n} -P_{\tilde U^n \tilde V^n}(u^n,v^n) \log \left(P_{U^nV^n}(u^n,v^n) \right) \notag \\
     &= \sum_{(u^n,v^n) \in \Ucal^n \times \V^n} \sum_{i=1}^n -P_{\tilde U^n \tilde V^n}(u^n,v^n) \log \left(P_{UV}(u_i,v_i) \right) \notag \\
   & = \sum_{i=1}^n \sum_{(u_i,v_i) \in \Ucal \times \V} -P_{\tilde U_i\tilde V_i}(u_i,v_i) \log \left(P_{UV}(u_i,v_i) \right) \notag \\
   &=\sum\nolimits_{i=1}^n H(\tilde U_i,\tilde V_i)+D(P_{\tilde U_i\tilde V_i}||P_{UV}) \notag  \\
   &=n \left(H(\tilde U_Q,\tilde V_Q)+D(P_{\tilde U_Q\tilde V_Q}||P_{UV})\right), \label{singletstp5}\\[5 pt]
   & H(\tilde V^n|\tilde U^n)+D(P_{\tilde U^n \tilde V^n}||P_{U^nV^n}) \notag  \\
   & =H(\tilde V^n,\tilde U^n)-H(\tilde U^n)+D(P_{\tilde U^n \tilde V^n}||P_{U^nV^n}) \notag \\
   & \geq H(\tilde V^n,\tilde U^n)+D(P_{\tilde U^n \tilde V^n}||P_{U^nV^n})-\sum\nolimits_{i=1}^n H(\tilde U_i ) \notag \\
   &= n \left( H(\tilde V_Q,\tilde U_Q)+D(P_{\tilde U_Q \tilde V_Q}||P_{UV})\right)-nH(\tilde U_Q) \label{applysingletuandv} \\
   & =n \left( H(\tilde V_Q|\tilde U_Q)+D(P_{\tilde U_Q \tilde V_Q}||P_{UV}) \right), \label{singletstp6} &&
\end{flalign}
where  we used \eqref{singletstp5} in \eqref{applysingletuandv}. Combining \eqref{singletstp1}, \eqref{singletstp3},  \eqref{singletstp2},  \eqref{singletstp5} and \eqref{singletstp6} yields \eqref{subaddit}.
\end{appendices}

\bibliographystyle{IEEEtran}
\bibliography{references2}
\end{document}